\documentclass[prb,aps,twocolumn,superscriptaddress,showpacs]{revtex4}

\usepackage{graphicx}% Include figure files

\begin{document}

\title{Evidence for decay of spin-waves above the pseudogap in underdoped YBa$_{2}$Cu$_{3}$O$_{6.35}$}

\author{C. Stock}
\affiliation{Department of Physics and Astronomy, Johns Hopkins University, Baltimore, Maryland USA 21218}

\author{R. A. Cowley}
\affiliation{Oxford Physics, Clarendon Laboratory, Parks Road, Oxford, United Kingdom OX1 3PU}

\author{W. J. L. Buyers}
\affiliation{National Research Council, Chalk River, Ontario, Canada K0J 1JO}
\affiliation{Canadian Institute of Advanced Research, Toronto, Ontario, Canada M5G 1Z8}

\author{R. Coldea}
\affiliation{Oxford Physics, Clarendon Laboratory, Parks Road, Oxford, United Kingdom OX1 3PU}

\author{C. L. Broholm}
\affiliation{Department of Physics and Astronomy, Johns Hopkins University, Baltimore, Maryland USA 21218}

\author{C.D. Frost}
\affiliation{Rutherford Appleton Laboratory, Chilton, Didcot, Oxon, United Kingdom OX11 0QX}

\author{R. J. Birgeneau}
\affiliation{Department of Physics, University of California at Berkeley, Berkeley, CA 94720}
\affiliation{Canadian Institute of Advanced Research, Toronto, Ontario, Canada M5G 1Z8}

\author{R. Liang}
\affiliation{Physics Department, University of British Columbia, Vancouver, B. C., Canada V6T 2E7}
\affiliation{Canadian Institute of Advanced Research, Toronto, Ontario, Canada M5G 1Z8}

\author{D. Bonn}
\affiliation{Physics Department, University of British Columbia, Vancouver, B. C., Canada V6T 2E7}
\affiliation{Canadian Institute of Advanced Research, Toronto, Ontario, Canada M5G 1Z8}

\author{W. N. Hardy}
\affiliation{Physics Department, University of British Columbia, Vancouver, B. C., Canada V6T 2E7}
\affiliation{Canadian Institute of Advanced Research, Toronto, Ontario, Canada M5G 1Z8}

\date{\today}

\begin{abstract}

	The magnetic spectrum at high-energies in heavily underdoped YBa$_{2}$Cu$_{3}$O$_{6.35}$ (T$_{c}$=18 K) has been determined throughout the Brillouin zone. At low-energy the scattering forms a cone of spin excitations emanating from the antiferromagnetic (0.5, 0.5) wave vector with an acoustic velocity similar to that of insulating cuprates. At high energy transfers, below the maximum energy of 270 meV at (0.5, 0), we observe zone boundary dispersion much larger and spectral weight loss more extensive than in insulating antiferromagnets.  Moreover we report phenomena not found in insulators, an overall lowering of the zone-boundary energies and a large damping of $\sim$ 100 meV of the spin excitations at high-energies. The energy above which the damping occurs coincides approximately with the gap determined from transport measurements.  We propose that as the energy is raised the spin excitations encounter an extra channel of decay into particle-hole pairs of a continuum that we associate with the pseudogap.
 
\end{abstract}


\maketitle

	There is a close relationship in high temperature superconductors between the carrier density, the magnetic ordering and excitations and the unusual $d$-wave pairing.~\cite{Lee06:78,Dai00:63,Fong00:61,Birgeneau06:75} The superconducting phase first appears when the concentration of carriers is about 5.5 \% and this small concentration of defects is also able to destroy the long range antiferromagnetic order. One of the most surprising results in the underdoped region of the phase diagram is the occurrence of a marked decrease in the charge scattering rate of the optical conductivity below the surmised pseudogap energy.~\cite{Timusk99:62} The pseudogap energy decreases with increasing doping and its origin is controversial with many models and theories proposed.~\cite{Stajic03:68,Anderson04:16,Ohkawa74:06,Franz87:01,Lee408:04} Understanding underdoped high temperature superconductors clearly requires an understanding of this pseudogap and more generally of how the antiferromagnetic Mott insulator becomes a superconductor for such small doping.

	YBa$_{2}$Cu$_{3}$O$_{6.35}$ (YBCO$_{6+x}$ with $x$=0.35) is a superconductor (T$_{c}$=18 K) with a planar hole doping of about 6\%. Low energy neutron scattering has shown that at this doping there is strong quasi-elastic scattering, showing the existence of commensurate antiferromagnetic short range order, and that there is no well-defined resonant mode.~\cite{Stock_unpub} In contrast in YBCO$_{6.5}$, and for similar hole doping, with T$_{c}$ = 59 K and a doping of 9 \%, the commensurate quasi-elastic peak is replaced by incommensurate excitations for energies less than a resonance at 33 meV where the excitations become commensurate at the antiferromagnetic zone centre (0.5, 0.5) in units of $2\pi/a$.~\cite{Stock05:71,Hayden04:429,Reznik04:93,Arai99:83,Tranquada04:429,Christensen04:93}  Since both these materials are superconducting it follows that none of these low energy properties of the magnetism are essential for the occurrence of superconductivity. Our paper represents the first observation by a direct spectroscopic probe of the destruction of spins by the pseudogap states, and has revealed an energy range where this onset occurs.  Based on a comparison to cuprates in the overdoped region of the phase diagram, we propose that this decay of spin excitations is a common feature in all cuprate high-temperature superconductors.

	To date there are very few measurements of the spin excitations at energies comparable with the zone boundary excitations at about 2$J$ in the parent antiferromagnetic insulators and none exist for the heavily underdoped superconducting materials.~\cite{Hayden96:54,Coldea01:86,Hayden96:76} We report neutron scattering measurements in this region for YBCO$_{6.35}$. We compare the results with those obtained for insulating two-dimensional antiferromagnets and show that the low energy magnetic excitations are very similar.  The high-energy excitations are very different and exhibit strong damping and reduced spectral weight near the magnetic zone boundary.  These properties are characteristic of the dispersion curve entering a continuum of excitations as found in magnetic metals and low dimensional materials.  We suggest that the contiuum is associated with the pseudogap states and the large damping occurs when the energy of the magnetic excitations is larger than the pseudogap.  

	The sample consisted of seven twinned crystals with a total volume of $\sim$ 7 cm$^{3}$ aligned with a combined rocking curve width of $\sim$ 2$^{o}$ and such that Bragg reflections (HHL) lay within the horizontal plane. The orthorhombic lattice parameters of the crystals were $a$=3.843 \AA\, $b$=3.871 \AA\ and $c$=11.788 \AA. The measurements were performed at high energies using the MAPS time-of-flight spectrometer at the ISIS facility.  A Fermi chopper was spun at 400 Hz and monochromated the incident beam with energies of 150 meV, 310 meV, and 450 meV. The effective resolution was selected by combining pixels and time channels using the MSLICE programme.~\cite{Coldea_mslice}  Typically the energy integration was chosen to be 10-30 meV (full-width at half maximum) while the integration in wave-vector transfer was between 0.025 and 0.1 reciprocal lattice units in the two-dimensional plane.  

\begin{figure}[t]
\includegraphics[width=7.5cm] {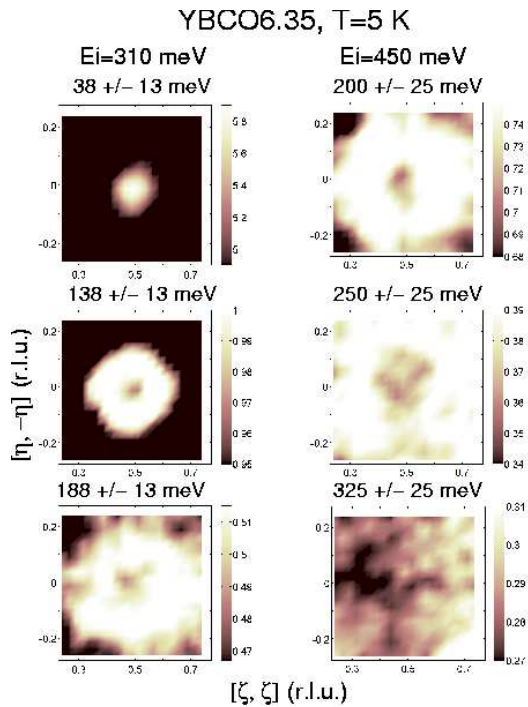}
\caption{\label{2dcuts}  Constant energy slices are shown.  The wavevector transfer is given by $(H,K) = (\zeta+\eta, \zeta-\eta)$.}
\end{figure}

\begin{figure}[t]
\includegraphics[width=7.5cm] {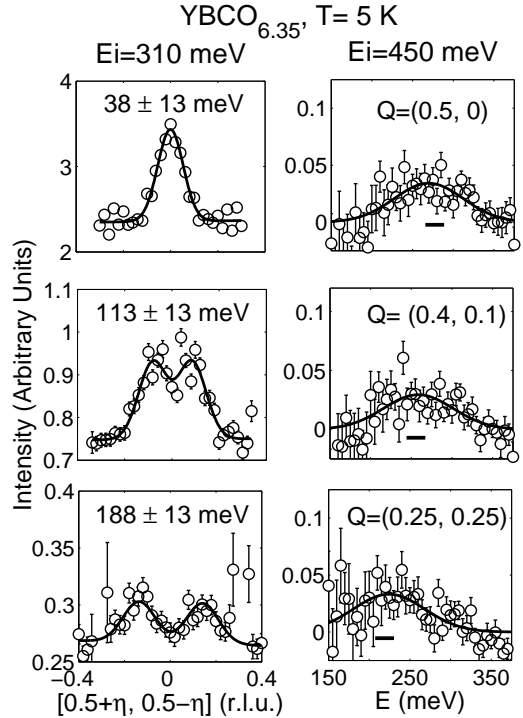}
\caption{\label{1dcuts} The left hand panels show cuts at constant energy transfer and with E$_{i}$=310 meV.  Constant-$Q$ cuts taken with E$_{i}$=450 meV are shown in the right hand panels for momentum positions around the zone boundary.  The data was integrated over ($\Delta$H, $\Delta$K)=($\pm$0.05, $\pm$0.05) and a background was taken at (0.5, 0.5).}
\end{figure}

\begin{figure}[t]
\includegraphics[width=8.0cm] {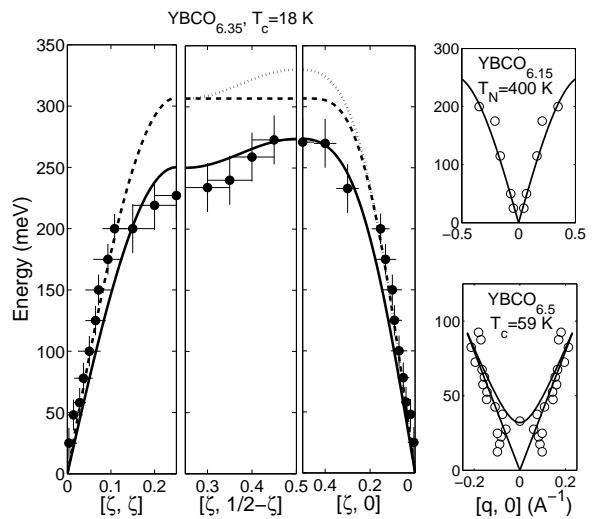}
\caption{\label{dispersion}  The momentum dispersion is compared with the insulator YBCO$_{6.15}$ (Ref. \onlinecite{Hayden96:54}) and YBCO$_{6.5}$ (Ref. \onlinecite{Stock05:71}).  $\zeta$ is measured from the zone centre (0.5, 0.5).}
\end{figure}

	In Fig. \ref{2dcuts} we show constant energy slices through the intensity distribution of magnetic neutron scattering for incident energies of 310 meV and 450 meV.  The coloured contour plots show how the scattering depends on wave-vector transfer in the two-dimensional CuO$_{2}$ plane.  In Fig. \ref{2dcuts}, the incident neutron beam was parallel to the c-axis and the results have been symmetrised to take account of the intensity arising at equivalent wave-vector transfers.  At low energies, the scattering is confined to a small circular disc centred on the antiferromagnetic Brillouin zone centre, (0.5, 0.5).  With increasing energy the size of the circle increases and the centre is depleted so as to form a ring whose diameter increases linearly with energy.  Above $\sim$ 200 meV the pattern becomes more complicated with the scattering intensity displaced towards the zone boundaries.  In Fig. \ref{1dcuts} we show a series of cuts through the data either to give constant energy scans or to give constant wave-vector scans through the data so as to obtain the dispersion from the antiferromagnetic peak position at (0.5, 0.5) as shown in Fig.\ref{dispersion}. 

\begin{figure}[t]
\includegraphics[width=7.5cm] {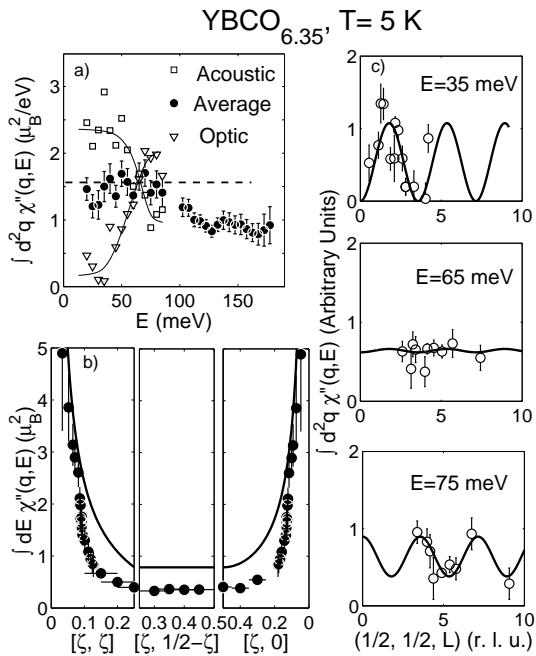}
\caption{\label{intensity} \textit{a)} The momentum integrated intensity is plotted as a function of energy transfer. \textit{b)} The energy integrated intensity as a function of wave vector relative to (0.5, 0.5) around the zone boundaries.  \textit{c)} The intensity as a function of L used to extract the acoustic and optic intensities.}
\end{figure}

	To extract the energy dependence of the acoustic and optic spectral weight, we varied the angle $\psi$, defined as the angle between the incident beam and the [001] axis, with the rotation axis along [1$\overline{1}$0].  With the incident energy fixed at 150 meV and angles of $\psi$=0, 20, and 30 the variation in the intensity with $L$ enables the optic and acoustic components to be separated by Fourier analysis knowing the bilayer spacing $d/c$=0.28.  This is the same analysis and method applied previously to the YBCO$_{6.5}$ superconductor.~\cite{Stock05:71}  As shown in Fig. \ref{intensity} $c)$ At low energies below 40 meV, all the scattering occurs in the acoustic channel.  At higher energies the optic mode intensity increases and at 75 meV it is larger than the acoustic mode intensity as shown by the reversal of the maxima and minima compared with the results at E=35 meV.  At larger energies the dispersion energies and intensities of both optic and acoustic modes are within error the same and similar to the insulating antiferromagnet. 

  	For energy transfers below 200 meV, the momentum integral of the scattered intensity is shown in Fig. \ref{intensity} $a)$ as a function of the two-dimensional wave vector.  Linear spin wave theory with a constant renormalization factor, Z$_{\chi}$=0.51 (Ref. \onlinecite{Igarashi92:46}), gives the dashed line.  The intensity loss at high energies while subject to uncertainties in the absolute normalization and in the estimation of the background is likely a forerunner of the large zone boundary loss discussed below.  For energy transfers below $\sim$ 200 meV, the magnetic excitation energies and intensities resemble the insulating antiferromagnet YBCO$_{6.15}$  and the acoustic and optic intensities are similar to those of the insulator. In particular, the slope of the dispersion curve gives a spin-wave velocity of 592 $\pm$ 30 meV $\cdot$ \AA, which is in agreement with the velocity of $\sim$ 600 meV $\cdot$ \AA\ measured for insulating YBCO$_{6.15}$.~\cite{Shamoto93:48} There are, however, considerable differences from spin wave behaviour at high-energy transfers above the range of $\hbar \omega \sim$ 150-200 meV.

	Firstly, the energy widths of the excitations, shown in the constant Q plots in Fig. \ref{1dcuts} are about 100 meV which are much larger than the experimental resolution (horizontal bar) of about 10 meV.  Therefore, the zone boundary excitations have acquired strong damping.

	Secondly, we observe an overall lowering of the zone boundary energies that is inconsistent with that of a simple antiferromagnet with a coupling between local spins. The dashed line in  Fig. \ref{dispersion} shows that a nearest neighbour model $J$=130 meV can fit the velocity at low-energies but fails to describe the dispersion at high energies above $\sim$ 150-200 meV. The dotted curve shows that the La$_{2}$CuO$_{4}$ nearest and next-nearest interactions of 112 meV and -11 meV respectively, also overestimates the frequencies of the zone boundary excitations.  The solid line represents an attempt to reproduce the zone boundary energies (with nearest and next nearest neighbour interactions of 96 and -10 meV respectively) and it shows that the data cannot be described by a consistent description with only near neighbour interactions.  All calculations include a renormalization factor calculated to be $Z_{c}$=1.18.~\cite{Igarashi92:46}  We also observe anisotropic behaviour at the zone boundary almost twice as large as that found in the insulator La$_{2}$CuO$_{4}$.  The zone boundary dispersion in YBCO$_{6.35}$ carries the same sign as that observed in La$_{2}$CuO$_{4}$ and opposite to that in the layered Sr$_{2}$Cu$_{3}$O$_{4}$Cl$_{2}$ and CFTD.~\cite{Kim99:83,Ronnow01:87} 

	A third difference between our results and linear spin wave theory is shown in Fig. \ref{intensity} $b)$ which plots the integral over all energies taken from constant-$Q$ cuts above 150 meV.  The zone boundary intensity is about a factor of two less than predicted by linear spin wave theory with a renormalization factor (Z$_{\chi}$=0.51).~\cite{Coldea01:86,Igarashi92:46}  Figure \ref{intensity} $b)$  also shows that the difference decreases on approaching the magnetic zone centre.  A decrease in the zone boundary magnetic neutron scattering intensity has also been suggested based on measurements in superconducting La$_{1.86}$Sr$_{0.14}$CuO$_{4}$.~\cite{Hayden96:76}  We note that in the square lattice CFTD, the spectral weight loss occurs only near the point (0.5, 0) whereas we observe a loss of spectral weight across a broader range in momentum.~\cite{Ronnow01:87} 

         The large energy widths, reduced zone boundary energy, and suppressed spectral weight suggest that the magnetic excitations interact with other high energy excitations that do not scatter neutrons strongly.  The magnetic excitations could decay into two spin-waves but we find that it is not possible to conserve wave vector and energy with the creation of pairs of uncoupled magnetic excitations having the dispersion of Fig. \ref{dispersion}.~\cite{Huberman05:72}  We note, however, that if the single-magnon linewidth is increased, then the kinematic constraints are relaxed and the decay into a pair of magnons may be possible.  It is also possible that in a S=1/2 system, there are two spin-wave states that are coupled and that lie below the two spin wave continuum. The interaction between these states and the magnetic excitations might give rise to the damping.  

	The fact that the unusual high-energy dispersion cannot be made consistent with the spin-wave models that describe the low-energy velocity and intensity (and also the entire spin-wave dispersion in the insulators) points to the need for an energy dependent decay process.  It is likely that the damping arises from the decay into electronic particle-hole pairs since the most energetic spin excitations lie above estimates for the pseudogap. One suggestion from thermal conductivity measurements is that the quasi-particle gap for heavily underdoped YBCO is $\sim$ 160 meV, which is comparable to the energy where the magnetic excitations show strong damping.~\cite{Sutherland05:94} The energy is also comparable to the expected pseudogap energy measured by NMR and optical absorption.~\cite{Timusk99:62}  Hence, for energy transfers greater than the pseudogap, the decay of a magnon into a particle-hole pair is allowed.  

	In ferromagnetic metallic systems characterized by a continuum of particle-hole excitations (for example in Fe, Ni, or MnSi), spin-waves become strongly damped when both $q$ and $\omega$ match those of the particle hole pairs.~\cite{Paul88:38,Ishikawa77:16}  This has been suggested to account for the anomalous decrease in intensity in YBCO$_{6.85}$ for energies $\sim$ 55 meV.~\cite{Pailhes04:93}  A similar decay process may account for strong damping of the high-energy excitations in YBCO$_{6.35}$.  It therefore appears that the large damping of the spin-excitations at high-energies maybe a common trait amongst all high-temperature superconductors across the entire phase diagram unlike the resonance and incommensurate scattering which have not been observed in all superconductors.  It would be interesting to model this effect however, the t-J model has been applied to heavily doped superconductors, but to our knowledge predictions for the lightly doped superconductor have not been made, possibly because of uncertainty in the Fermi surface.~\cite{Kao00:61,Norman:01:63,Bulut93:47,Schnyder05:unpub,Chen05:71,Bascones05:unpub}
	
	Near the onset of superconductivity, the low energy spin response is strong and gives the appearance of spin-waves while at high energy we have shown that the spin excitations enter a region where the decay, decrease in energy, and intensity do not fit into a spin-wave description.  The evidence points to the presence of a particle-hole continuum as a possible explanation of the results.  On this basis, we consider that the pseudogap is identified to be in the range 150-200 meV for $x$=0.35, T$_{c}$=18 K.  The damping observed for $x$=0.85, T$_{c}$=85 K also suggests that the spin excitations decay near 55 meV and if this has the same origin it implies that the pseudogap has decreased with increasing doping.  In contrast, the low energy results between $x$=0.35 and $x$=0.5 (Fig. \ref{dispersion}) show the presence or absence of quasi-elastic scattering, commensurate or incommensurate excitations, and the presence or absence of a resonance suggesting that these features are not present in all superconductors.  We therefore suggest that the existence of the high energy continuum seems to be the feature which is common to all superconductors at least up to optimal doping, whereas the low energy features seem less universal.  

We are grateful for the assistance of the staff at ISIS and for financial support from the Natural Science and Engineering Research Council (NSERC) of Canada and the US National Science Foundation through DMR-0306940.

\end{document}